\magnification 1200
% Psfig/TeX Release 1.2
% dvips version
%
% All software, documentation, and related files in this distribution of
% psfig/tex are Copyright 1987, 1988 Trevor J. Darrell
%
% Permission is granted for use and non-profit distribution of psfig/tex 
% providing that this notice be clearly maintained, but the right to
% distribute any portion of psfig/tex for profit or as part of any commercial
% product is specifically reserved for the author.
%
% $Header: psfig.tex,v 1.9 88/01/08 17:42:01 trevor Exp $
% $Source: $
%
% Thanks to Greg Hager (GDH) and Ned Batchelder for their contributions
% to this project.
%
\catcode`\@=11\relax
\newwrite\@unused
\def\typeout#1{{\let\protect\string\immediate\write\@unused{#1}}}
\typeout{psfig/tex 1.2-dvips}

%% Here's how you define your figure path.  Should be set up with null
%% default and a user useable definition.

\def\figurepath{./}

%
% @psdo control structure -- similar to Latex @for.
% I redefined these with different names so that psfig can
% be used with TeX as well as LaTeX, and so that it will not 
% be vunerable to future changes in LaTeX's internal
% control structure,
%
\def\@nnil{\@nil}
\def\@empty{}
\def\@psdonoop#1\@@#2#3{}
\def\@psdo#1:=#2\do#3{\edef\@psdotmp{#2}\ifx\@psdotmp\@empty \else
    \expandafter\@psdoloop#2,\@nil,\@nil\@@#1{#3}\fi}
\def\@psdoloop#1,#2,#3\@@#4#5{\def#4{#1}\ifx #4\@nnil \else
       #5\def#4{#2}\ifx #4\@nnil \else#5\@ipsdoloop #3\@@#4{#5}\fi\fi}
\def\@ipsdoloop#1,#2\@@#3#4{\def#3{#1}\ifx #3\@nnil 
       \let\@nextwhile=\@psdonoop \else
      #4\relax\let\@nextwhile=\@ipsdoloop\fi\@nextwhile#2\@@#3{#4}}
\def\@tpsdo#1:=#2\do#3{\xdef\@psdotmp{#2}\ifx\@psdotmp\@empty \else
    \@tpsdoloop#2\@nil\@nil\@@#1{#3}\fi}
\def\@tpsdoloop#1#2\@@#3#4{\def#3{#1}\ifx #3\@nnil 
       \let\@nextwhile=\@psdonoop \else
      #4\relax\let\@nextwhile=\@tpsdoloop\fi\@nextwhile#2\@@#3{#4}}
\def\psdraft{
	\def\@psdraft{0}
	%\typeout{draft level now is \@psdraft \space . }
}
\def\psfull{
	\def\@psdraft{100}
	%\typeout{draft level now is \@psdraft \space . }
}
\psfull
\newif\if@prologfile
\newif\if@postlogfile
\newif\if@noisy
\def\pssilent{
	\@noisyfalse
}
\def\psnoisy{
	\@noisytrue
}
\psnoisy
%%% These are for the option list.
%%% A specification of the form a = b maps to calling \@p@@sa{b}
\newif\if@bbllx
\newif\if@bblly
\newif\if@bburx
\newif\if@bbury
\newif\if@height
\newif\if@width
\newif\if@rheight
\newif\if@rwidth
\newif\if@clip
\newif\if@verbose
\def\@p@@sclip#1{\@cliptrue}

%%% GDH 7/26/87 -- changed so that it first looks in the local directory,
%%% then in a specified global directory for the ps file.

\def\@p@@sfile#1{\def\@p@sfile{null}%
	        \openin1=#1
		\ifeof1\closein1%
		       \openin1=\figurepath#1
			\ifeof1\typeout{Error, File #1 not found}
			\else\closein1
			    \edef\@p@sfile{\figurepath#1}%
                        \fi%
		 \else\closein1%
		       \def\@p@sfile{#1}%
		 \fi}
\def\@p@@sfigure#1{\def\@p@sfile{null}%
	        \openin1=#1
		\ifeof1\closein1%
		       \openin1=\figurepath#1
			\ifeof1\typeout{Error, File #1 not found}
			\else\closein1
			    \def\@p@sfile{\figurepath#1}%
                        \fi%
		 \else\closein1%
		       \def\@p@sfile{#1}%
		 \fi}

\def\@p@@sbbllx#1{
		%\typeout{bbllx is #1}
		\@bbllxtrue
		\dimen100=#1
		\edef\@p@sbbllx{\number\dimen100}
}
\def\@p@@sbblly#1{
		%\typeout{bblly is #1}
		\@bbllytrue
		\dimen100=#1
		\edef\@p@sbblly{\number\dimen100}
}
\def\@p@@sbburx#1{
		%\typeout{bburx is #1}
		\@bburxtrue
		\dimen100=#1
		\edef\@p@sbburx{\number\dimen100}
}
\def\@p@@sbbury#1{
		%\typeout{bbury is #1}
		\@bburytrue
		\dimen100=#1
		\edef\@p@sbbury{\number\dimen100}
}
\def\@p@@sheight#1{
		\@heighttrue
		\dimen100=#1
   		\edef\@p@sheight{\number\dimen100}
		%\typeout{Height is \@p@sheight}
}
\def\@p@@swidth#1{
		%\typeout{Width is #1}
		\@widthtrue
		\dimen100=#1
		\edef\@p@swidth{\number\dimen100}
}
\def\@p@@srheight#1{
		%\typeout{Reserved height is #1}
		\@rheighttrue
		\dimen100=#1
		\edef\@p@srheight{\number\dimen100}
}
\def\@p@@srwidth#1{
		%\typeout{Reserved width is #1}
		\@rwidthtrue
		\dimen100=#1
		\edef\@p@srwidth{\number\dimen100}
}
\def\@p@@ssilent#1{ 
		\@verbosefalse
}
\def\@p@@sprolog#1{\@prologfiletrue\def\@prologfileval{#1}}
\def\@p@@spostlog#1{\@postlogfiletrue\def\@postlogfileval{#1}}
\def\@cs@name#1{\csname #1\endcsname}
\def\@setparms#1=#2,{\@cs@name{@p@@s#1}{#2}}
%
% initialize the defaults (size the size of the figure)
%
\def\ps@init@parms{
		\@bbllxfalse \@bbllyfalse
		\@bburxfalse \@bburyfalse
		\@heightfalse \@widthfalse
		\@rheightfalse \@rwidthfalse
		\def\@p@sbbllx{}\def\@p@sbblly{}
		\def\@p@sbburx{}\def\@p@sbbury{}
		\def\@p@sheight{}\def\@p@swidth{}
		\def\@p@srheight{}\def\@p@srwidth{}
		\def\@p@sfile{}
		\def\@p@scost{10}
		\def\@sc{}
		\@prologfilefalse
		\@postlogfilefalse
		\@clipfalse
		\if@noisy
			\@verbosetrue
		\else
			\@verbosefalse
		\fi
}
%
% Go through the options setting things up.
%
\def\parse@ps@parms#1{
	 	\@psdo\@psfiga:=#1\do
		   {\expandafter\@setparms\@psfiga,}}
%
% Compute bb height and width
%
\newif\ifno@bb
\newif\ifnot@eof
\newread\ps@stream
\def\bb@missing{
	\if@verbose{
		\typeout{psfig: searching \@p@sfile \space  for bounding box}
	}\fi
	\openin\ps@stream=\@p@sfile
	\no@bbtrue
	\not@eoftrue
	\catcode`\%=12
	\loop
		\read\ps@stream to \line@in
		\global\toks200=\expandafter{\line@in}
		\ifeof\ps@stream \not@eoffalse \fi
		%\typeout{ looking at :: \the\toks200 }
		\@bbtest{\toks200}
		\if@bbmatch\not@eoffalse\expandafter\bb@cull\the\toks200\fi
	\ifnot@eof \repeat
	\catcode`\%=14
}	
\catcode`\%=12
\newif\if@bbmatch
\def\@bbtest#1{\expandafter\@a@\the#1%%BoundingBox:\@bbtest\@a@}
\long\def\@a@#1%%BoundingBox:#2#3\@a@{
\ifx\@bbtest#2\@bbmatchfalse\else\@bbmatchtrue\fi}
\long\def\bb@cull#1 #2 #3 #4 #5 {
	\dimen100=#2 bp\edef\@p@sbbllx{\number\dimen100}
	\dimen100=#3 bp\edef\@p@sbblly{\number\dimen100}
	\dimen100=#4 bp\edef\@p@sbburx{\number\dimen100}
	\dimen100=#5 bp\edef\@p@sbbury{\number\dimen100}
	\no@bbfalse
}
\catcode`\%=14
\def\compute@bb{
		\no@bbfalse
		\if@bbllx \else \no@bbtrue \fi
		\if@bblly \else \no@bbtrue \fi
		\if@bburx \else \no@bbtrue \fi
		\if@bbury \else \no@bbtrue \fi
		\ifno@bb \bb@missing \fi
		\ifno@bb \typeout{FATAL ERROR: no bb supplied or found}
			\no-bb-error
		\fi
		\count203=\@p@sbburx
		\count204=\@p@sbbury
		\advance\count203 by -\@p@sbbllx
		\advance\count204 by -\@p@sbblly
		\edef\@bbw{\number\count203}
		\edef\@bbh{\number\count204}
		%\typeout{ bbh = \@bbh, bbw = \@bbw }
}
%
% \in@hundreds performs #1 * (#2 / #3) correct to the hundreds,
%	then leaves the result in @result
%
\def\in@hundreds#1#2#3{\count240=#2 \count241=#3
		     \count100=\count240	% 100 is first digit #2/#3
		     \divide\count100 by \count241
		     \count101=\count100
		     \multiply\count101 by \count241
		     \advance\count240 by -\count101
		     \multiply\count240 by 10
		     \count101=\count240	%101 is second digit of #2/#3
		     \divide\count101 by \count241
		     \count102=\count101
		     \multiply\count102 by \count241
		     \advance\count240 by -\count102
		     \multiply\count240 by 10
		     \count102=\count240	% 102 is the third digit
		     \divide\count102 by \count241
		     \count200=#1\count205=0
		     \count201=\count200
			\multiply\count201 by \count100
		 	\advance\count205 by \count201
		     \count201=\count200
			\divide\count201 by 10
			\multiply\count201 by \count101
			\advance\count205 by \count201
		     \count201=\count200
			\divide\count201 by 100
			\multiply\count201 by \count102
			\advance\count205 by \count201
		     \edef\@result{\number\count205}
}
\def\compute@wfromh{
		% computing : width = height * (bbw / bbh)
		\in@hundreds{\@p@sheight}{\@bbw}{\@bbh}
		%\typeout{ \@p@sheight * \@bbw / \@bbh, = \@result }
		\edef\@p@swidth{\@result}
		%\typeout{w from h: width is \@p@swidth}
}
\def\compute@hfromw{
		% computing : height = width * (bbh / bbw)
		\in@hundreds{\@p@swidth}{\@bbh}{\@bbw}
		%\typeout{ \@p@swidth * \@bbh / \@bbw = \@result }
		\edef\@p@sheight{\@result}
		%\typeout{h from w : height is \@p@sheight}
}
\def\compute@handw{
		\if@height 
			\if@width
			\else
				\compute@wfromh
			\fi
		\else 
			\if@width
				\compute@hfromw
			\else
				\edef\@p@sheight{\@bbh}
				\edef\@p@swidth{\@bbw}
			\fi
		\fi
}
\def\compute@resv{
		\if@rheight \else \edef\@p@srheight{\@p@sheight} \fi
		\if@rwidth \else \edef\@p@srwidth{\@p@swidth} \fi
}
%		
% Compute any missing values
\def\compute@sizes{
	\compute@bb
	\compute@handw
	\compute@resv
}
%
% \psfig
% usage : \psfig{file=, height=, width=, bbllx=, bblly=, bburx=, bbury=,
%			rheight=, rwidth=, clip=}
%
% "clip=" is a switch and takes no value, but the `=' must be present.
\def\psfig#1{\vbox {
	% do a zero width hard space so that a single
	% \psfig in a centering enviornment will behave nicely
	%{\setbox0=\hbox{\ }\ \hskip-\wd0}
	%
	\ps@init@parms
	\parse@ps@parms{#1}
	\compute@sizes
	\ifnum\@p@scost<\@psdraft{
		\if@verbose{
			\typeout{psfig: including \@p@sfile \space }
		}\fi
		\special{ps::[begin] 	\@p@swidth \space \@p@sheight \space
				\@p@sbbllx \space \@p@sbblly \space
				\@p@sbburx \space \@p@sbbury \space
				startTexFig \space }
		\if@clip{
			\if@verbose{
				\typeout{(clip)}
			}\fi
			\special{ps:: doclip \space }
		}\fi
		\if@prologfile
		    \special{ps: plotfile \@prologfileval \space } \fi
		\special{ps: plotfile \@p@sfile \space }
		\if@postlogfile
		    \special{ps: plotfile \@postlogfileval \space } \fi
		\special{ps::[end] endTexFig \space }
		% Create the vbox to reserve the space for the figure
		\vbox to \@p@srheight true sp{
			\hbox to \@p@srwidth true sp{
				\hss
			}
		\vss
		}
	}\else{
		% draft figure, just reserve the space and print the
		% path name.
		\vbox to \@p@srheight true sp{
		\vss
			\hbox to \@p@srwidth true sp{
				\hss
				\if@verbose{
					\@p@sfile
				}\fi
				\hss
			}
		\vss
		}
	}\fi
}}

\catcode`\@=12\relax
% usage : \psfig{file=, height=, width=, bbllx=, bblly=, bburx=, bbury=,
%			rheight=, rwidth=, clip=}
\hsize 16. true cm             
% \hoffset xyz                 
% \vsize 25. true cm                                     
\def\baseaa{\baselineskip = 4truemm}        
\def\basea{\baselineskip = 6truemm}        
\def\baseb{\baselineskip = 6truemm}        
\basea
% \nopagenumbers                                         
%
%%%%%%%%%%%%%%%%%%%%%%%%%%%%%%%%%%%%%%%%%%%%%%%%%%%%%%%%%%%%%%%%%%%
%%%%%%%%%%%%%%%%%%%%%%%%%%%%%%%%%%%%%%%%%%%%%%%%%%%%%%%%%%%%%%%%%%%
\def\Z#1{\zeta(#1)} 
\def\parn{\par\noindent}
\def\app{{\left({\alpha\over \pi}\right)}}
\def\ref#1{[#1]}

\def\d${$ \displaystyle }

\def\o{\omega}

\def\A#1{a_{#1}}

\def\ll{\ln{\lambda}}

\def\om{\omega}
\def\gm#1{$(g$-$#1)$} 
%%%%%%%%%%%%%%%%%%%%%%%%%%%%%%%%%%%%%%%%%%%%%%%%%%%%%%%%%%%%%%%%%%%
%%%%%%%%%%%%%%%%%%%%%%%%%%%%%%%%%%%%%%%%%%%%%%%%%%%%%%%%%%%%%%%%%%%
\basea
\rightline{\bf DFUB 96-05}
\rightline{7 February 1996}  
\vskip 12truemm           
\centerline{\bf The analytical value of the electron 
                ($g$-$2$) at order $\alpha^3$ in QED. } 
\par 
\vskip 12truemm                                                  
%%%%%%%%%%%%%%%%%%%%%%% footnote %%%%%%%%%%%%%%%%%%%%%%%%%%%%%%%%%%%%%%%
\centerline{S.Laporta
\footnote {$^{\S}$}{ {E-mail: {\tt laporta@bo.infn.it} }}  }
\centerline{E.Remiddi
\footnote {$^{\circ}$}{ {E-mail: {\tt remiddi@bo.infn.it} }}  }
%%%%%%%%%%%%%%%%%%%%%%% footnote %%%%%%%%%%%%%%%%%%%%%%%%%%%%%%%%%%%%%%%
\vskip 12truemm                                                  
\centerline{\it Dipartimento di Fisica, Universit\`a di Bologna,}
\centerline{\it and INFN, Sezione di Bologna,}
\centerline{\it Via Irnerio 46, I-40126 Bologna, Italy}
\vskip 12truemm                                                  
%ABSTRACT-++++++++++++++++++++++++++++++++++++++++++++++++++++++++++
\centerline{\bf Abstract }  \par                                
We have evaluated in closed analytical form the contribution 
of the three-loop non-planar `triple-cross' diagrams contributing to the 
electron ($g$-$2$) in QED; its value, omitting the already known 
infrared divergent part, is 
$$ \eqalign{ 
      a_e({\rm 3-cross}) = 
  & \phantom{+} {1\over 2} \pi^2 \Z3 - {55\over 12} \Z5 
    - {16\over 135} \pi^4
    + {32\over 3} \left(\A4 + {1\over 24} {\ln^4 2}\right) \cr
  & + {14\over 9} \pi^2 {\ln^2 2} 
    - {1\over 3} \Z3 + {23\over 3} \pi^2 {\ln 2}
    - {47\over 9} \pi^2  - {113\over 48}  \ . \cr }  $$ 
This completes the analytical evaluation of the \gm2 at order 
$\alpha^3$, giving 
$$ \eqalign{
 a_e({\rm 3-loop}) = \app^3 \biggl\{ 
 & \phantom{+} {83\over 72}  \pi^2 \Z3 - {215\over 24} \Z5 
   + {100\over 3} \left[\left( \A4 + {1\over24} {\ln^4 2} \right) 
                        - {1\over24} \pi^2 {\ln^2 2}  \right] \cr
 & - {239\over 2160} \pi^4 
   + {139\over 18} \Z3  - {298\over 9}  \pi^2 {\ln 2}
   + {17101\over 810} \pi^2 + {28259\over 5184} \biggr\} \cr =  
 &  \app^3 (1.181241456...) \ .\cr  } $$
%ABSTRACT-++++++++++++++++++++++++++++++++++++++++++++++++++++++++++ 
% \vskip 6truemm                                                  
\phantom{la vispa teresa} \parn 
\baseaa 
\baseaa 
\item{{\it PACS:}} 12.20Ds; 13.40.Em; 06.20.Jr; 12.20Fv
\item{{\it Keywords:}} Quantum ElectroDynamics; Anomalous magnetic moment of 
                the electron; Analytical evaluation of 3-loop radiative 
                corrections. 
%\centerline {--------------------------------------------------------------} 
\vfill\eject 
%%%%%%%%%%%%%%%%%%%%%%%%%%%%%%%%%%%%%%%%%%%%%%%%%%%%%%%%%%%%%%%%%% 
% qui ci si potrebbe mettere il testo 
\baseb
\baseb
Following the work of Ref. [1] we have completed the evaluation in close 
analytical form of the contribution to the electron anomaly in three-loop 
QED due to the triple-cross graphs depicted in Fig. 1). 
%%%%%%%%%%%%%%%%%%%%%%%%%%%%%%%%%%%%%%%%%%%%%%%%%%%%%%%%%%%%%%%%%%%%%%%%%
%%%%%%%%%%%%%%%%%%%%%%%%%%%%%%%%%%%%%%%%%%%%%%%%%%%%%%%%%%%%%%%%%%%
%%%%%%%%%%%%%%%%%%%%%%%%%%%%%%%%%%%%%%%%%%%%%%%%%%%%%%%%%%%%%%%%%%%
%%%%%%%%%%%%%%%%%%%%%%%%%%%%%%%%%%%%%%%%%%%%%%%%%%%%%%%%%%%%%%%%%%%
\vskip -3.0truecm 
\psfig{figure=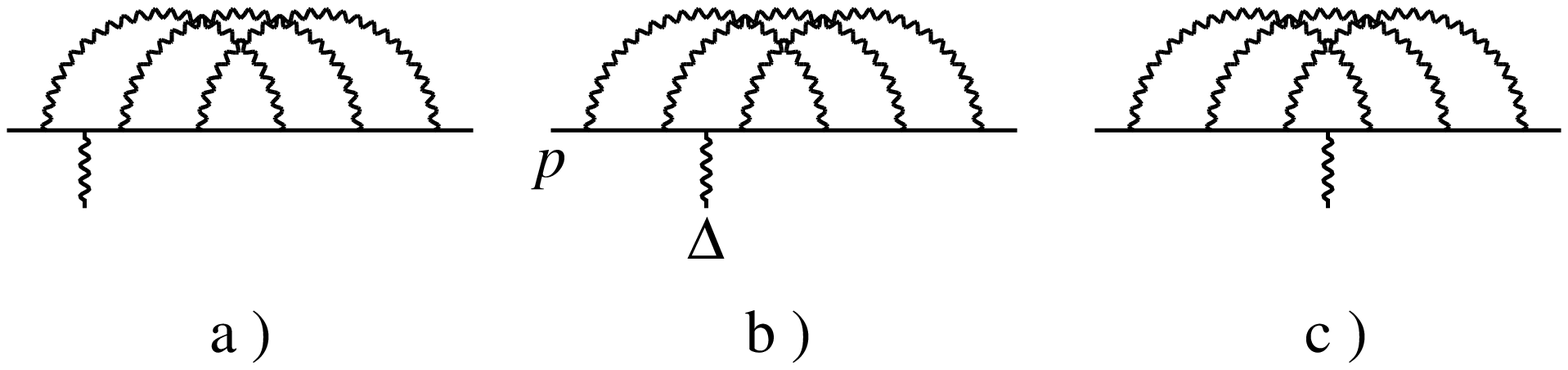,height=20cm} 
\vskip -10.5truecm 
\centerline{ Fig.1. The ``triple-cross" graphs. } 
\vskip 0.5truecm 
%%%%%%%%%%%%%%%%%%%%%%%%%%%%%%%%%%%%%%%%%%%%%%%%%%%%%%%%%%%%%%%%%%%
%%%%%%%%%%%%%%%%%%%%%%%%%%%%%%%%%%%%%%%%%%%%%%%%%%%%%%%%%%%%%%%%%%%%%%%%%
The results are 
$$ \eqalign{ 
 a_e({\rm 3-cross}; 1a) = 
& \phantom{+} {215\over 24} \Z5 - {1\over 3} \pi^2 \Z3 
  - {53\over 2160} \pi^4
  + 4\left[\left( \A4 + {1\over24} {\ln^4 2}\right) 
                      - {1\over24} \pi^2 {\ln^2 2} \right] \cr
& - {1229\over 144} \Z3 - {7\over 6} \pi^2 {\ln 2}  + {4165\over 2592} \pi^2
  - {515\over 864}  + {1\over2} {\ln\lambda} \cr =
& \ 1.285068495... + {1\over2} {\ln\lambda} \ ,\cr } \eqno(1)$$ 
$$ \eqalign{ 
  a_e({\rm 3-cross}; 1b) = 
 & - {275\over 24} \Z5    + {29\over 36} \pi^2 \Z3
   + {43\over 1080} \pi^4  - {5\over 6} \pi^2 {\ln^2 2} 
   + {623\over 144} \Z3   + {35\over 9} \pi^2 {\ln 2}  \cr
 & - {1951\over 648} \pi^2   - {493\over 864} \cr  =
 & -0.878968171... \ ,\cr }  \eqno(2)$$
$$ \eqalign{
  a_e({\rm 3-cross}; 1c) = 
  & \phantom{+} {5\over 12} \Z5    - {4\over 9} \pi^2 \Z3
    - {161\over 1080} \pi^4
    + {8\over 3} \left(\A4 +{1\over 24} {\ln^4 2}\right) 
    + {32\over 9} \pi^2 {\ln^2 2} \cr
  &  + {97\over 12} \Z3   + {20\over 9} \pi^2 {\ln 2}
    - {1043\over 432} \pi^2  - {1\over 48} \cr  =
  & -0.026799490... \ ,\cr } \eqno(3) $$
from which 
$$ \eqalign{ 
      a_e({\rm 3-cross}) \ =& 
    \ 2 a_e({\rm 3-cross};1a) + 2 a_e({\rm 3-cross};1b) 
    + a_e({\rm 3-cross};1c) \cr 
  =& \phantom{+} {1\over 2} \pi^2 \Z3 - {55\over 12} \Z5 
    - {16\over 135} \pi^4
    + {32\over 3} \left(\A4 + {1\over 24} {\ln^4 2}\right) \cr
   & + {14\over 9} \pi^2 {\ln^2 2} 
    - {1\over 3} \Z3 + {23\over 3} \pi^2 {\ln 2}
    - {47\over 9} \pi^2  - {113\over 48} + {\ln\lambda} \cr  
  =& \; 0.785401156... + {\ln\lambda} \ . \cr }  \eqno(4) $$ 
As it is customary in this kind of calculations, 
$\lambda$ is the regularizing photon mass (in electron mass units), 
$ \zeta(p) =\sum\limits_{n=1}^{\infty}{1/n^p} ,$ 
$ \Z2=\pi^2/6$, 
\hbox{ \d$ a_4 =\sum\limits_{n=1}^{\infty}{1\over{2^nn^4}} $\ . } 
\par 
The above results are in excellent agreement with the numerical
results
$$ \eqalign { a_e({\rm 3-cross}; 1a; Ref.2) &= \phantom{-}1.291(7)\ , \cr 
              a_e({\rm 3-cross}; 1b; Ref.2) &= -0.882(10)\ , \cr 
              a_e({\rm 3-cross}; 1c; Ref.2) &= -0.021(100) \ , \cr 
              a_e({\rm 3-cross}; Ref.3)     &= \phantom{-}0.785419(40)\ , } $$ 
where the $ {\ln\lambda}$'s are understood. 
As the previous graphs were the last graphs for which the analytical 
value of the anomaly was still missing, on account the previously 
known results [4,5,6,7] the complete analytical expression of the 
anomaly in three-loop QED can now be written as 
$$ \eqalign{
 a_e({\rm 3-loop}) = \app^3 \biggl\{ 
 & \phantom{+} {83\over 72}  \pi^2 \Z3 - {215\over 24} \Z5 
   + {100\over 3} \left[\left( \A4 + {1\over24} {\ln^4 2} \right) 
                        - {1\over24} \pi^2 {\ln^2 2}  \right] \cr
 & - {239\over 2160} \pi^4 
   + {139\over 18} \Z3  - {298\over 9}  \pi^2 {\ln 2}
   + {17101\over 810} \pi^2 + {28259\over 5184} \biggr\} \cr =  
 &  \app^3 (1.181241456...) \ .\cr  } \eqno(5) $$
By using the best numerical value of $a_e({\rm 4-loop})= -1.557(70)\ , $ 
Ref.[8], and 
$$ 1/\alpha= 137.0359979(32) \ , $$ 
Ref.[9], one finds 
$$ a_e({\rm th}) = 1159652201.2(2.1)(27.1) \times 10^{-12}\ , \eqno(6)$$ 
to be compared with the experimental value, Ref.[10], 
$$ a_e({\rm exp}) = 1159652188.4(4.3) \times 10^{-12}\ ; $$ 
conversely, by using Ref.[10] as an input, one obtains 
$$ 1/\alpha(a_e) = 137.03599941(56)\ . \eqno(7) $$ 
\vskip 1.0 truecm 
\def\ieps{ i\epsilon } 
If $p$ is the electron momentum and $\Delta$ the momentum transfer 
of some vertex amplitude, the corresponding \gm2 contribution is extracted 
by keeping $p$ on the mass shell, expanding the vertex up to first order 
in $\Delta$, multiplying it by a suitable spinor projection operator 
and then performing the appropriate traces [11]. 
The \gm2 is then expressed as the sum a few hundred terms all of the kind 
$$ \int d^4k_1 d^4k_2 d^4k_3  {N\over D} \ , \eqno(8)$$ 
each term having its own specific numerator $N$ and denominator $D$. 
The numerators $N$ are in general simple monomials in the scalar products 
of $p$ and the loop momenta $k_i$, such as for instance 
$ (p\cdot k_i),\ (k_i\cdot k_j)\ $ etc, up to $(p\cdot k_2)^4, $ 
while the denominators $ D $ have in general the form 
% $$ D= \DPKae^{n_1}\DPKabe^{n_2}\DPKabce^{n_3}\DPKbce^{n_4}
%       \DPKce^{n_5}\DKae^{n_6}\DKbe^{n_7}\DKce^{n_8} $$
$$ D= D_1^{n_1} D_2^{n_2} D_3^{n_3} D_4^{n_4}
      D_5^{n_5} D_6^{n_6} D_7^{n_7} D_8^{n_8} \ , $$
the exponents $n_i$ ranging from 0 to 2 (0 means that the denominator 
is absent; the powers 2 arise from the expansion in $\Delta$; 
in each term there are at most two denominators with $n_i=2$); 
in $m_e=1$ units the denominators, see Fig.2), are defined as 
$$ \eqalign{ 
   D_1 =& (p-k_1)^2+1-\ieps\ , \hskip 19mm D_2 = (p-k_1-k_2)^2+1-\ieps\ , \cr 
   D_3 =& (p-k_1-k_2-k_3)^2+1-\ieps\ , \quad D_4 = (p-k_2-k_3)^2+1-\ieps\ , \cr 
   D_5 =& (p-k_3)^2+1-\ieps\ , \quad D_6 = k_1^2-\ieps\ , \quad 
   D_7 =  k_2^2-\ieps\ ,       \quad D_8 = k_3^2-\ieps \ . } \eqno(9)$$ 
Besides the 8 denominators, there are in the problem 9 linearly independent 
scalar products $p\cdot k_i, k_i\cdot k_j,$ etc; 8 independent identities 
between the scalar products and the denominators are easily written, 
such as for instance 
$$ \eqalign{
  p  \cdot k_1 =& {1\over2}(D_6 - D_1) \ , \cr
  k_1\cdot k_3 =& {1\over2}(- D_2 + D_3 - D_4 + D_7) - {p\cdot k_2} \ . 
                                 } \eqno (10)$$
The identities can be used to express the scalar products in the 
numerator as a combination of the denominators, so obtaining a combination 
of integrands with a smaller number of denominators and therefore 
simpler from the point of view of the analytical integration. 
It is to observed however that, as there are 9 scalar products
and only 8 denominators, one scalar product (which we choose to be 
$p\cdot k_2$) must remain anyhow in the numerator; furthermore, the 
simpler terms obtained in that way can be in general non convergent 
when taken separately . 
\parn 
%%%%%%%%%%%%%%%%%%%%%%%%%%%%%%%%%%%%%%%%%%%%%%%%%%%%%%%%%%%%%%%%%%%%%%%%%%%%%%%%
%%%%%%%%%%%%%%%%%%%%%%%%%%%%%%%%%%%%%%%%%%%%%%%%%%%%%%%%%%%%%%%%%%%%%%%%%%%%%%%%
%%%%%%%%%%%%%%%%%%%%%%%%%%%%%%%%%%%%%%%%%%%%%%%%%%%%%%%%%%%%%%%%%%%
\basea
\vskip 1.3truecm 
\hskip 2.5truecm 
\psfig{figure=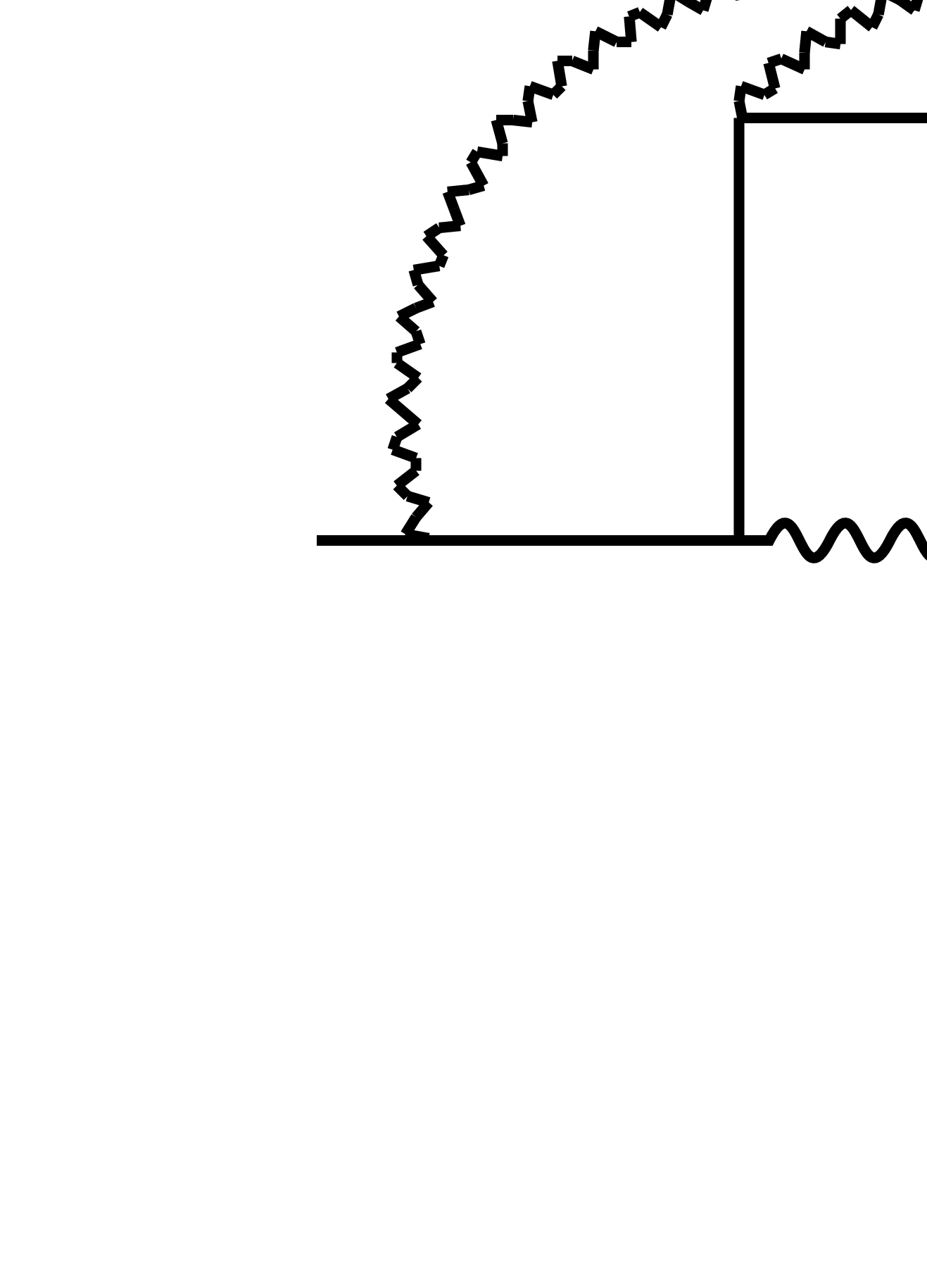,height=5cm} 
%%%%%%%%%%%%%%%%%%%%%% PERICOLO ! %%%%%%%%%%%%%%%%%%%%%%%%%%%%%%%%%%
%%% NON cambiare alla leggera i \vskip wtc. (non commutano) 
%%% regola [?] con \vkip negativo  si risale rispetto al punto precedente 
\vskip -3.2truecm \hskip 3.5truecm $p$ \hskip 6.6truecm $p$ 
\vskip -1.9truecm \hskip 4.7truecm $D_6$ \hskip 3.6truecm $D_8$ 
\vskip  0.5truecm \hskip 5.0truecm $D_1$ \hskip 1.2truecm $D_7$ 
                  \hskip 1.1truecm $D_5$ 
\vskip -1.3truecm \hskip 6.1truecm $D_2$ \hskip 0.7truecm $D_4$ 
\vskip -1.4truecm \hskip 6.9truecm $D_3$ 
\vskip  1.55truecm \hskip 6.9truecm $k_2$ 
\vskip -3.25truecm \hskip 4.5truecm $k_1$ \hskip 4.2truecm $k_3$ 
%%% regola [?] con \vkip positivo  si scende rispetto al bottom 
\vskip 3truecm 
\centerline{ Fig.2. The notation for the 3-loop integrations.} 
\vskip 0.5truecm 
%%%%%%%%%%%%%%%%%%%%%%%%%%%%%%%%%%%%%%%%%%%%%%%%%%%%%%%%%%%%%%%%%%%%%%%%%%%%%%%%
\baseb
%%%%%%%%%%%%%%%%%%%%%%%%%%%%%%%%%%%%%%%%%%%%%%%%%%%%%%%%%%%%%%%%%%%%%%%%%%%%%%%%
In Ref. [1] we gave the analytical value of the simplest term 
$$ J_0 \equiv {(-i)^3 \over\pi^6}\int d^4k_1 d^4k_2 d^4k_3\ 
                 {1\over{D_1 D_2 D_3 D_4 D_5 D_6 D_7 D_8}}  = 
              4\pi^2\ln^2 2 -{1\over6}\pi^4 \ .                  \eqno(11) $$
%             24 \Z2 \ln^2 2 -6\zeta^2(2) \ .                    \eqno(47) $$
With a straightforward extension of the results of Ref. [1] one can obtain, 
for instance 
$$ J_1 \equiv {(-i)^3 \over\pi^6} \int d^4k_1 d^4k_2 d^4k_3\ 
          {{p\cdot k_2}\over{D_1 D_2 D_3 D_4 D_5 D_6 D_7 D_8}}  = 
           5\Z5 - {1\over2}\pi^2\Z3 \ .                          \eqno(12) $$ 
The extension to all the other terms appearing in the expression giving 
the \gm2 is in principle straightforward but in fact rather wearisome 
(it implies, among other things, to build, for each tensor in the loop 
momenta, the scalar amplitudes free from kinematical singularities for 
which unsubtracted dispersion relations can be written). 
It was found convenient to work out another approach, based on the 
integration by part identity method first developed in Ref.[12]. 
In our case the method consists in writing 
$$ \int \left(\prod\limits_{i=1}^3d^{n}k_i\right) 
              {\partial\over\partial k_{j,\mu}} 
             \left( v_{l,\mu}{N\over D} \right) = 0 \ , \eqno(13) $$ 
where $v_l$ is either the external momentum $p$ or one of the three loop 
momenta $k_i$, $N/D$ stands for the ratio of all the pairs of numerators 
and denominators of the kind described above; note however that while 
in the original integrals the loop momenta have dimension 4, the 
identities are written for momenta having continuous $n$ dimension, 
in which case their validity is obvious. 
The required $n=4$ values are then recovered by means of a suitable 
limiting procedure. 
Needless to say, if the \gm2 from some graph is obtained as the 
$n\to 4$ limit of an $n$-dependent expression, this expression must be 
evaluated for consistency in $n$ dimensions from the very beginning; 
in particular, the traces of the $\gamma$-matrix must be evaluated 
in $n$-dimensions too. 
\par 
By explicitly performing the derivatives, 
using the identities (10)
and rearranging the obtained 
terms in an identity like Eq.(13), one obtains a relation between 
various integrals which 
amounts to express one of them as a linear combination of the others. 
The number of all the possible identities is large (a few thousands, as 
there are twelve identities for each of the several hundreds $N/D$ 
possibilities); they form a huge linear system, whose coefficients are 
in turn polynomials in $n$, from which the overwhelming majority of the 
occurring integrals can be expressed as a linear combination of a few basic 
integrals. As we are interested in the $n\to4$ limit, we put 
$n=4-2\o$ and then deal with the limit by expanding everything 
in powers of $\o$ for $\o\to0$ . 
Fortunately, the expansion and the extremely lengthy processing of the 
various identities can be carried out in an almost mechanical way. 
\par 
\def\norm{\left({{-i}\over{\pi^{n-2}}}\right)^3
                                    \int d^nk_1 d^nk_2 d^nk_3\ } 
A thorough investigation shows that all the three loop integrals 
(and therefore also the complete \gm2 expression in which we are 
interested) can be expressed in terms of the following 18 basic integrals: 
$$ I_{1} =\norm {p\cdot k_2 \over D_1 D_2 D_3 D_4 D_5 D_6 D_7 D_8} \ ,$$ %%% NZ
$$ I_{2}   =\norm {1\over D_1 D_2 D_3 D_4 D_7 D_8} \ ,$$ %%% I6f
$$ I_{3}   =\norm {1\over D_1 D_2 D_4 D_5 D_6 D_8} \ ,$$ %%% I6e 
$$ I_{4}   =\norm {1\over D_2 D_3 D_4 D_6 D_7 D_8} \ ,$$ %%% I6d
$$ I_{5}   =\norm {1\over D_1 D_3 D_4 D_5 D_7 D_8} \ ,$$ %%% I6c
$$ I_{6}   =\norm {1\over D_1 D_3 D_5 D_6 D_7 D_8} \ ,$$ %%% I6b
$$ I_{7}   =\norm {1\over D_2 D_4 D_5 D_6 D_7 D_8} \ ,$$ %%% I6a
$$ I_{8}   =\norm {1\over D_1 D_2 D_3 D_4 D_5} \ ,$$ %%% I5c
$$ I_{9}   =\norm {1\over D_2 D_3 D_5 D_6 D_7} \ ,$$ %%% I5b
$$ I_{10}  =\norm {1\over D_2 D_4 D_6 D_7 D_8} \ ,$$ %%% I5a
$$ I_{11}  =\norm {p\cdot k_2 \over D_1 D_3 D_5 D_7} \ ,$$ %%%=I4f1
$$ I_{12}  =\norm {1\over D_1 D_3 D_5 D_7} \ ,$$ %%% I4f
$$ I_{13}  =\norm {1\over D_1 D_2 D_4 D_5} \ ,$$ %%% I4e
$$ I_{14}  =\norm {1\over D_3 D_5 D_6 D_7} \ ,$$ %%% I4d
$$ I_{15}  =\norm {1\over D_2 D_3 D_4 D_5} \ ,$$ %%% I4c
$$ I_{16}  =\norm {1\over D_3 D_4 D_7 D_8} \ ,$$ %%% I4b
$$ I_{17}  =\norm {1\over D_3 D_6 D_7 D_8} \ ,$$ %%% I4a
$$ I_{18}  =\norm {1\over D_1 D_4 D_5} \ .$$  %%% I3a
Note that in the above list there is only one integral with all the 8 
denominators ($I_1$, the limit of which is Eq.(12) above), while there are no 
integrals with 7 denominators. With the exception of $I_1$, all the basic 
integral are divergent in the $n\to4$ limit; having put $n=4-2\omega$, 
the divergences show up as $1/\omega$ singularities, so that evaluating 
them amounts to evaluate the coefficients of their expansions in 
$\omega$. 
As an example, the value of $ I_{18} $, which perhaps the easiest to 
obtain as it factorizes in the product of three 1-loop integrals, is 
$$ I_{18} = C(\o)\left( -{1\over\om^3} -{3\over\om^2} -{6\over\om} - 10
     -15\om -21\om^2 - 28\om^3 + O(\om^4) \right) \ , \eqno(14) $$ 
where $ C(\om) $, defined as 
$$ C(\om) = \left(\pi^{\om}\Gamma(1+\om)\right)^3 \ ,            $$ 
is an overall normalization factor, whose limiting value at $\omega=0$ 
is 1. 
\par 
As an example of the kind of formulae obtained by expressing 
a non-trivial integral in terms of the basic integrals, let us give 
the relation 
$$ \eqalign{
  {(-i)^3 \over\pi^6} &\int d^4k_1 d^4k_2 d^4k_3\
  {(p\cdot k_2)^2 \over D_1 D_2 D_3 D_4 D_5 D_6 D_7 D_8} = 
         \lim_{\o\to0} \biggl[ \quad   - I_1             \cr
%%%   &+ I_2 \left(  - 3\o - 8\o^2 - 24\o^3 \right) 
   &+ I_2 \left(  - 3\o  \right) 
    + I_3 \left( {1\over4} - {3\over4}\o - 2\o^2 - 6\o^3 \right) \cr 
%%%   &+ I_4 \left( {3\over2}\o + {7\o^2} + {32\o^3} \right)
   &+ I_4 \left( {3\over2}\o \right)
    + I_5  \left(  - \o - 2\o^2 - 4\o^3 \right)
    + I_6  \left( {1\over2} - {1\over2}\o - 3\o^2 - 16\o^3 \right) \cr 
   &+ I_7  \left( 3\o + 14\o^2 + 64\o^3 \right)
    + I_8  \left(  - {1\over6\o}\right)
    + I_9  \left(  - {1\over2\o} - {1\over2} + \o + 4\o^2 + 176\o^3 \right) \cr 
   &+ I_{10} \left(  - {1\over2} - {5\over2}\o - {27\over2}\o^2 
                  - {137\over2}\o^3 \right)                            \cr 
   &+ I_{11} \left(   {1\over6\o} - {139\over36} - {1249\over108}\o 
                  - {19225\over324}\o^2 - {238333\over972}\o^3 \right) \cr 
   &+ I_{12} \left(  {3\over64\o^2} +{13\over128\o} +{383\over128} 
                 + 2\o - {353\over16}\o^2 + {3041\over8}\o^3 \right)   \cr 
   &+ I_{13} \left(  -{1\over12\o} - {3\over8} - {1\over6}\o 
                   +{29\over24}\o^2 + {37\over4}\o^3 \right)           \cr 
   &+ I_{14} \left( -{1\over4\o} +{13\over8} + {87\over8}\o 
                  +{439\over8}\o^2 + {623\over2}\o^3 \right)           \cr 
   &+ I_{15} \left( -{7\over12\o} -{11\over8} - {19\over24}\o 
                  +{55\over12}\o^2 +{73\over2}\o^3 \right)             \cr 
   &+ I_{16} \left(  -{1\over2\o} -{7\over4} -{7\over4}\o +{9\over4}\o^2 
                  + 196\o^3 \right)                                    \cr 
   &+ I_{17} \left( {3\over2\o} +{3\over4} +{29\over6}\o +{50\over3}\o^2 
%%%                  -{2698\over3}\o^3 \right)                            \cr 
                  \right)                            \cr 
   &+ I_{18} \left(   {3\over64\o^2} -{2\over3\o} -{55\over36} 
             -{1391\over864}\o -{94423\over5184}\o^2
             +{3535507\over7776}\o^3 \right) \biggr] \ .    \cr }   
                                                     \eqno(15) $$
Similar expressions hold for all the other integrals contributing 
to the anomaly. In all those expressions, the coefficient of $I_1$ is 
always finite ({\it i.e.} not singular for $\omega\to 0$), so that 
only its $\o=0$ limit, given by $J_1$ of Eq.(12), is required. 
\par 
Besides $I_1$, we have already given the explicit analytical value of 
$I_{18}$, Eq.(14). In order to complete the calculation of all the 
basic integrals, we have evaluated directly only a subset of them 
(namely $I_8, I_9$, $I_{10}$ and from $I_{13}$ to $I_{18}$ ). 
For the others, we found it more convenient 
to write a large number of identities of the form of Eq.(15), 
in which however the analytical value of the integral on the 
{\it l.h.s.} is already known from previous work; the relation is 
then expressing such a known result in terms of the still unknown 
basic integrals. 
In the case of $J_0$, Eq.(11), one finds for instance 
$$ \eqalign{
  J_{0}& = \lim_{\o\to0} \biggl[ \phantom{+} 
          I_{2}   \left( {5\over 2} - {15\over 2} \om \right) 
        + I_{3}   \left( {3\over 4} - {9\over 4} \om \right) 
        + I_{4}   \left(  - {5\over 4} + {5\over 4} \om \right) \cr
    &   + I_{5}   \left( {1\over 2} - {3\over 2} \om \right) 
        + I_{6}   \left( {3\over 4} - {3\over 4} \om + {3\over 2} \om^2
                         + 6 \om^3 \right) 
        + I_{7}   \left(  - {5\over 2} + {5\over 2} \om - 5 \om^2
                          - 20 \om^3 \right) \cr
    &   + I_{8}   \left( - {2\over 3\om} +{4\over 3} - {16\over 3} \om
                         - {64\over 3} \om^2 - 128 \om^3 \right) 
        + I_{9}   \left(  {1\over 8\om^2} - {1\over\om}+ 1 + \om \right) \cr
    &   + I_{10}   \left( {1\over 4\om} + 1 + {7\over 2} \om
                        + {39\over 2} \om^2 + {201\over 2} \om^3 \right) \cr
    &   + I_{11}   \left(   {7\over 6\om}  - {49\over 36}
                           + {11\over 108} \om - {1819\over 324} \om^2
                           - {34249\over 972} \om^3\right) \cr
    &   + I_{12}   \left(- {3\over 128\om^3} + {59\over 256\om^2}
                         - {203\over 256\om} + {15\over 8} - {37\over 16} \om 
                         -{53\over 16} \om^2 - {113\over 8} \om^3 \right) \cr
    &   + I_{13}   \left(- {1\over 3\om}   - {17\over 24} - {211\over 48} \om
                       - {423\over 16} \om^2 - {1263\over 8} \om^3 \right) \cr
    &   + I_{14}   \left( {1\over 16\om^2} - {49\over 32\om} +{53\over 32}
                         - {3\over 8} \om + {119\over 16} \om^2
                         + {427\over 8} \om^3 \right) \cr
    &   + I_{15}   \left( - {4\over 3\om}  - {17\over 6} - {211\over 12} \om
                         - {423\over 4} \om^2 - {1263\over 2} \om^3 \right) \cr
    &   + I_{16}   \left( {1\over 8\om^2} - {13\over 16\om} +{7\over 16}
                         - {5\over 4} \om - {9\over 2} \om^2
                         - {67\over 4} \om^3 \right) \cr
    &   + I_{17}   \left(  - {3\over 4\om^2} + {23\over 8\om}  - {69\over 8}
                           - {40\over 3} \om - {587\over 6} \om^2 \right) \cr
    &   + I_{18}   \left(  - {3\over 128\om^3} + {7\over 32\om^2}
                           - {29\over 12\om}  - {3799\over 576} 
                          - {140051\over 3456} \om
                          - {1193035\over 5184} \om^2 
%%%&   \phantom{+I_{18}+}\left.
                          - {10388627\over 7776} \om^3 \right)\biggr].
                                                             } \eqno(16)$$
Similarly, in Ref.[7] the following 7-denominator integral 
was evaluated 
$$ {(-i)^3 \over\pi^6}\int d^4k_1 d^4k_2 d^4k_3\ 
                 {1\over{D_1 D_3 D_4 D_5 D_6 D_7 D_8}}  = 
             {21\over12}\pi^2\Z3 - {45\over4}\Z5 \ ;   \eqno(17) $$
%            {21\over2}\Z2\Z3 - {45\over4}\Z5 \ ;   \eqno(272) $$
in terms of the basic integrals one has 
$$ \eqalign{
% & \norm {1 \over D_1 D_3 D_4 D_5 D_6 D_7 D_8} = \cr 
& {(-i)^3 \over\pi^6}\int d^4k_1 d^4k_2 d^4k_3\ 
        {1 \over D_1 D_3 D_4 D_5 D_6 D_7 D_8} = \lim_{\o\to0} \biggl[ 
          I_{5}   \left(  {1\over 2\om}  - {3\over 2} \right)         \cr 
    &   + I_{6}   \left( - {1\over 2\om} +{1\over 2}- \om - 4 \om^2
                         - 16 \om^3 \right) 
        + I_{9}   \left( - {1\over 4\om}  +{1\over 2} \right) \cr
    &   + I_{11}   \left(  {2 \over \om} +{14\over 3} + {445\over 18} \om
                          + {3104\over 27} \om^2 + {85177\over 162} \om^3
                          \right) 
        + I_{12}   \left(  - {3\over 16} - {19\over 32} \om
                          - {39\over 16} \om^2 - {79\over 8} \om^3 \right) \cr
    &   + I_{14}   \left( - {11\over 8\om}  - {53\over 16} - {299\over 16} \om 
                          - {715\over 8} \om^2 - {1667\over 4} \om^3 \right) \cr
    &   + I_{15}   \left( - {1\over 4\om} + {1\over 8} - {1\over 8} \om
                          - {1\over 4} \om^2 - {1\over 2} \om^3 \right) 
        + I_{16}   \left(  {1\over 4\om} - {1\over 8} + {1\over 8} \om
                          + {1\over 4} \om^2 + {1\over 2} \om^3 \right) \cr
    &   + I_{17}   \left( - {1\over 2\om}  - {7\over 12} - {10\over 3} \om
%%% - {40\over 3} \om^2 - {160\over 3} \om^3 \right) \cr
                          - {40\over 3} \om^2 \right) \cr
    &   + I_{18}   \left(  {3\over 8\om} +{29\over 48} + {521\over 144} \om
                          + {7691\over 432} \om^2
                          + {111629\over 1296} \om^3 \right) \biggr]\ .\cr }
   \eqno(18) $$ 
It is to be stressed here that such relations have quite a broad generality, 
which extends to cover scalar amplitudes occurring in graphs with different 
topology; the {\it l.h.s.} integral of Eq.s(17),(18) was indeed 
evaluated in Ref.[7], dealing with ``corner-ladder'' graphs. 
\par 
By exploiting our previous work, we can write a redundant number of such 
relations expressing unknown basic integrals in terms of already known 
integrals (they are not written here for the sake of brevity); in so 
doing we obtain a redundant set of linear equations, which are easily 
solved in terms of the unknown basic integrals (the redundancy provides 
with additional consistency checks). As a side remark, let us observe 
that for each basic integral one must evaluate the first few 
terms (up to 7) of its expansion in $\o$, see for instance $I_{18}$ , 
so that the number of unknown quantities increases, but on the other 
hand each relation, being an identity in $\o$ provides with several 
independent equations for the unknown. 
\par 
As a result the following integral table is established 
$$  I_1=   C(\o)\biggl[   5 \Z5    -{1\over2}\pi^2 \Z3 +O(\o) \biggr] \ , $$
% $$  I_1=      5 \Z5    -3 \Z2 \Z3 \ , $$
$$ \eqalign{
 I_2=C(\o)\biggl[&\phantom{+} 2 {\Z3 \over \om}
       - {13\over 90} \pi^4 - {1\over 3} \pi^2 + 10 \Z3 \cr 
%%% modificato
   &+ \om \biggl( {385\over2} \Z5 -{85\over6} \pi^2 \Z3 -{7\over15} \pi^4 
                 -{82 \Z3 } -{4 \pi^2 {\ln 2}} +{16} \pi^2 
                  \biggr) +O(\o^2) \biggr] \ ,\cr } $$
$$ \eqalign{
  I_3=C(\o)\biggl[
   & {1\over 3\om^3} +  {7\over 3\om^2} +  {31\over 3\om} 
      - {2\over 15} \pi^4 - {4\over 3} \Z3 + {103\over 3} 
   + \om \biggl( 95\Z5 - {25\over 3}\pi^2\Z3 \cr&\  - {1\over 15}\pi^4 
              - {184\over 3} \Z3 - 8 \pi^2{\ln 2} + {44\over 3}\pi^2 
              + {235\over 3} \biggr) +O(\o^2)  \biggr]   \ , \cr } $$
$$ \eqalign{ I_4 = C(\o)\biggl[
   & 2 {\Z3\over \om} - {7\over 90} \pi^4 + 2 \Z3 + {1\over 3} \pi^2 \cr 
%%% modificato
   &+ \om \biggl( {385\over2} \Z5 -{85\over6} \pi^2 \Z3 -{7\over15} \pi^4 
                 -{82 \Z3 } -{4 \pi^2 {\ln 2}} +{16} \pi^2 
                   \biggr) +O(\o^2) \biggr] \ , \cr} $$
$$ \eqalign{
  I_5= C(\o)\biggl[
       & {1\over 6\om^3} + {3\over 2\om^2}
       + {1\over\om} \biggl(- {1\over 3} \pi^2 + {55\over 6} \biggr)
       - {4\over 45} \pi^4 - {14\over 3} \Z3 - {7\over 3} \pi^2 
                                             + {95\over 2} \cr 
      & + \om   \biggl( - {2\over 9} \pi^4 - 44 \Z3 - {29\over 3} \pi^2 
                         + {1351\over 6}  \biggr) +O(\o^2)\biggr]   \ ,\cr} $$
$$ \eqalign{
 I_6=C(\o)\biggl[
     &  {1\over 3\om^3} 
       + {7\over 3\om^2} 
       + {31\over 3\om} 
    - {4\over 45} \pi^4  + {2\over 3} \Z3 + {1\over 3} \pi^2 + {103\over 3}     
     + \om   \biggl(
       {45\over 2} \Z5 - {7\over 2} \pi^2\Z3 \cr& \ 
   + {11\over 45} \pi^4 + {14\over 3} \Z3  - 4 \pi^2{\ln 2} + {14\over 3} \pi^2 
     + {235\over 3}   \biggr) +O(\o^2)\biggr]  \ ,   \cr} $$
$$ \eqalign{
 I_7=C(\o)\biggl[
     &   {1\over 6\om^3}
       + {3\over 2\om^2}
       + {1\over \om} \biggl( - {1\over 3} \pi^2 + {55\over 6}  \biggr)
       - {1\over 15} \pi^4 - {8\over 3} \Z3 - 2 \pi^2 + {95\over 2}  
      + \om   \biggl(
     {45\over 2} \Z5 \cr&\  - {17\over 6} \pi^2\Z3 
   - {7\over 9} \pi^4 - 50 \Z3 - 4 \pi^2{\ln 2} + {1\over 3} \pi^2 
   +{1351\over 6}  \biggr) +O(\o^2)\biggr]    \ ,   \cr}$$
$$ \eqalign{
 I_8=C(\o)\biggl[&
       - {1\over \om^3}
       - {16\over 3\om^2}
       - {16\over \om}
       + 2 \Z3 - {8\over 3} \pi^2 - 20 \cr
     &  + \om \biggl(
      - {3\over 10} \pi^4  - {200\over 3} \Z3 + 16 \pi^2{\ln 2}
      - 28 \pi^2 + {364\over 3}    \biggr) \cr
%%% modificato
     & + \om^2   \biggl(
-{126 \Z5} +{21 \pi^2 \Z3} +{46\over15} \pi^4 -{512 \A4} -{64\over3} 
{\ln^4 2} \cr
     & \phantom{+\om^2\biggl(}
        -{80\over3} \pi^2 {\ln^2 2} -{776 \Z3} +{168 \pi^2 {\ln 2}} 
        -{188 \pi^2} +{1244} \biggr) +O(\o^3) \biggr] \ , \cr}$$
$$ \eqalign{
  I_9=C(\o)\biggl[&
       - {2\over 3\om^3}
       - {10\over 3\om^2}
       + {1\over\om} \biggl(- {1\over 3} \pi^2  - {26\over 3}  \biggr)
        - {16\over 3} \Z3 - {11\over 3} \pi^2 - 2 \cr
     & + \om   \biggl(- {13\over 45} \pi^4  - {248\over 3} \Z3 
       + 16 \pi^2{\ln 2} - {73\over 3} \pi^2  +{398\over 3}\biggr) \cr
%%% modificato
     & + \om^2   \biggl(
        -{96 \Z5} -{8\over3} \pi^2 \Z3 +{3\over5} \pi^4 -{512 \A4} 
        -{64\over3} {\ln^4 2} \cr
     & \phantom{+\om^2\biggl(}
       -{128\over3} \pi^2 {\ln^2 2} -{1888\over3} \Z3 +{160 \pi^2 {\ln 2}} 
       -{129 \pi^2} +{1038} \biggr) +O(\o^3)\biggr]  \ ,\cr }$$
$$ \eqalign{
I_{10}=C(\o)\biggl[
        & - {1\over 3\om^3} - {5\over 3\om^2} 
          + {1\over\om} \biggl( - {2\over 3} \pi^2  - 4  \biggr)
          - {26\over 3}\Z3 - {7\over 3} \pi^2 + {10\over 3} \cr 
        & + \om  \biggl(- {35\over 18} \pi^4  - {94\over 3} \Z3 
- \pi^2 +{302\over 3} \biggr)   +O(\o^2)\biggr]    \ ,  \cr} $$ 
$$ \eqalign{
 I_{11}=C(\o)\biggl[
     & {1\over 2\om^3} + {37\over 24\om^2} + {43\over 16\om}
       + 2 \Z3 + {139\over 96} 
       + \om\biggl( - {1\over 10} \pi^4 + {19\over 3} \Z3 
                    - {773\over 64}  \biggr) \cr 
%%% modificato
      & + \om^2\biggl( -{447\over2} \Z5 +{53\over2} \pi^2 \Z3 
                     -{67\over60} \pi^4 +{235\over2} \Z3 \cr
      & \phantom{+\om^2\biggl(}
+{12 \pi^2 {\ln 2}} -{18 \pi^2} -{27869\over384} \biggr) +O(\o^3) \biggr]
 \ ,   \cr   } $$ 
$$ \eqalign{
I_{12}=C(\o)\biggl[
  &  {1\over\om^3} + {7\over 2\om^2} + {253\over 36\om} + {2501\over 216}
     + \om  \biggl(- {64\over 9} \pi^2 + {59437\over 1296}  \biggr) \cr
    &+ \om^2\biggl( - {1792\over 9} \Z3 + {256\over 3} \pi^2{\ln 2} 
                      - {2272\over 27} \pi^2 +{2831381\over 7776}) \cr 
%%% modificato
    &+ \om^3\biggl( {2752\over135} \pi^4 -{8192\over3} \A4
 -{1024\over9} {\ln^4 2}
                    -{3584\over9} \pi^2 {\ln^2 2} \cr
    & \phantom{+\om^3\biggl(}
      -{63616\over27} \Z3 +{9088\over9} \pi^2 {\ln 2} -{49840\over81} \pi^2 
        +{117529021\over46656} \biggr)  +O(\o^4) \biggr]  \ ,  \cr}$$
$$ \eqalign{ 
I_{13}=C(\o)\biggl[
    & {2\over\om^3} + {23\over 3\om^2} + {35\over 2\om} + {275\over 12}
     + \om  \biggl({112\over 3} \Z3  - {189\over 8}  \biggr) 
+ \om^2\biggl( - {136\over 45} \pi^4 \cr &\ + 256 \A4 + {32\over 3} {\ln^4 2} 
                    - {32\over 3} \pi^2 {\ln^2 2}  + 280 \Z3  
                    - {14917\over 48}\biggr)  +O(\o^3)\biggr]    \ ,\cr}$$
$$ \eqalign{
 I_{14}=C(\o)\biggl[
    & {1\over 3\om^3} + {7\over 6\om^2} + {25\over 12\om} 
                      + {8\over 3} \Z3 - {5\over 24} \cr
    &+ \om \biggl( - {2\over 15} \pi^4 + {28\over 3} \Z3 
                   - {959\over 48}   \biggr) \cr
    &+ \om^2 \biggl( {48 \Z5} -{7\over15} \pi^4 +{50\over3} \Z3 
                   -{10493\over96} \biggr) +O(\o^3)\biggr]  \ , \cr} $$
$$ \eqalign{
 I_{15}=C(\o)\biggl[
      & {3\over 2\om^3} + {23\over 4\om^2} + {105\over 8\om}
                          + {4\over 3} \pi^2 + {275\over 16} 
     + \om  \biggl(  28 \Z3 - 8 \pi^2{\ln 2} + 10 \pi^2 
                    - {567\over 32} \biggr) \cr
%%% modificato
    &+ \om^2\biggl( - {62\over 45} \pi^4 + 192 \A4 + 8 {\ln^4 2} 
                    + 16 \pi^2 {\ln^2 2} + 210 \Z3 - 60 \pi^2{\ln 2} \cr
    &\phantom{+\om^2\biggl(}
+ {145\over 3} \pi^2  - {14917\over 64} \biggr) +O(\o^3) \biggr] \ , } $$
$$ \eqalign{
 I_{16}=C(\o)\biggl[
     & {1\over 2\om^3} + {7\over 4\om^2} 
     + {1\over\om} \biggl( {1\over 3} \pi^2 +{25\over 8}  \biggr)
     + 4 \Z3 + {7\over 6} \pi^2 - {5\over 16} \cr
    &+ \om \biggl(   { 16\over 45} \pi^4 + 14 \Z3 + {25\over 12} \pi^2 
                   - {959\over 32} \biggr) \cr
%%% modificato
    &+ \om^2 \biggl({72 \Z5} +{8\over3} \pi^2 \Z3 +{56\over45} \pi^4 +{25 \Z3} 
-{5\over24} \pi^2 -{10493\over64} \biggr)  +O(\o^3) \biggr] \ , }$$
$$ \eqalign{
I_{17}=C(\o)\biggl[
      & - {1\over 6\om^2} - {35\over 36\om} - {1\over 3} \pi^2 
         - {559\over 216} 
     + \om \biggl( - {16\over 3} \Z3 - {35\over 18} \pi^2 
                   + {2737\over 1296} \biggr) \cr
    &+ \om^2\biggl( - {37\over 45} \pi^4 - {280\over 9} \Z3 
- {559\over 108} \pi^2 +{552041\over 7776} \biggr) +O(\o^3) \biggr] \ .} $$
\par 
By using the above table in combination with the integration by parts 
identities, we obtain the analytical value of the required integrals; 
from Eq.(15) we have for instance 
$$\eqalign{
   {(-i)^3 \over\pi^6} \int d^4k_1 d^4k_2 d^4k_3\
  &  {(p\cdot k_2)^2 \over D_1 D_2 D_3 D_4 D_5 D_6 D_7 D_8} = \cr 
  &  = {1\over2} \Z3 \pi^2 - 5 \Z5 - {2\over45}\pi^4 
     + \Z3  - 2 \pi^2\ln2 + {3\over2}\pi^2 \ .  \cr}  \eqno(18)$$ 
\par 
In terms of the basic integrals, the \gm2 of the graph of Fig.1c), for 
instance, reads 
$$ \eqalign{
 &(g-2)({\rm Fig.1c)}) = \lim_{\o\to0} \biggl[ 
  I_{1} \biggl({7\over6} \biggr) 
 +I_{2} \biggl( -{1\over2\o}  +{19\over12} -{1637\over72}\o  \biggr) \cr
&+I_{3} \biggl( -{13\over48\o} +{253\over144} -{823\over108}\o 
                +{102797\over2592}\o^2 
                -{979525\over3888}\o^3 \biggr) 
 +I_{4} \biggl(  {3\over8\o}   -{5\over24} +{229\over18}\o  \biggr) \cr
&+I_{5} \biggl( -{1\over12\o} -{7\over18} -{587\over216}\o 
                +{9133\over324}\o^2 -{340685\over1944}\o^3 \biggr) \cr
&+I_{6} \biggl( -{1\over24\o} +{55\over72} -{827\over108}\o 
                +{19075\over648}\o^2 -{874721\over3888}\o^3 \biggr) \cr
&+I_{7} \biggl(  {19\over24\o} -{25\over18} +{632\over27}\o 
                -{56983\over648}\o^2 +{1488295\over1944}\o^3 \biggr) \cr
&+I_{8} \biggl(  {1\over8\o^2} -{5\over9\o} +{1585\over216} 
                -{36581\over1296}\o +{1051253\over3888}\o^2 
                -{743606\over729}\o^3 \biggr) \cr
&+I_{9} \biggl(  {11\over96\o^2} -{565\over288\o} +{467\over36} 
                -{11225\over162}\o +{1629641\over3888}\o^2 
                -{1068433\over432}\o^3 \biggr) \cr
&+I_{10} \biggl(-{11\over72\o} -{613\over216} -{187\over324}\o 
                -{234293\over3888}\o^2 
                +{1054147\over23328}\o^3 \biggr) \cr
&+I_{11} \biggl(-{1\over6\o^2} -{1\over18\o} -{2143\over216}
                +{5767\over144}\o -{2472821\over7776}\o^2 
                +{20840363\over11664}\o^3 \biggr) \cr
&+I_{12} \biggl(-{11\over512\o^3} +{525\over1024\o^2} -{6601\over2304\o}
                +{577597\over27648} -{4981223\over41472}\o 
                +{46026487\over62208}\o^2  \cr
&\phantom{+I_{12}\biggl(}\ 
                -{1632171647\over373248}\o^3 \biggr) \cr
&+I_{13} \biggl( {7\over96\o^2} +{13\over192\o} +{2375\over576} 
                -{95\over432}\o +{339883\over2592}\o^2
                +{60055\over3888}\o^3 \biggr) \cr
&+I_{14} \biggl( {35\over192\o^2} -{1051\over1152\o} +{29419\over1728}
                -{220349\over3456}\o +{264997\over486}\o^2 
                -{31472155\over11664}\o^3 \biggr) \cr
&+I_{15} \biggl( {1\over4\o^2} +{7\over36\o} +{6229\over432} 
                +{187\over27}\o +{1934333\over3888}\o^2
                +{1580447\over11664}\o^3 \biggr) \cr
&+I_{16} \biggl( {11\over96\o^2} -{1075\over576\o} +{8179\over864}
                -{10925\over192}\o +{2919977\over7776}\o^2 
                -{49397413\over23328}\o^3 \biggr) \cr
&+I_{17} \biggl(-{17\over24\o^2} +{1429\over144\o} -{12223\over216}
                +{146749\over432}\o 
                -{15117413\over7776}\o^2  \biggr) \cr
&+I_{18} \biggl(-{11\over512\o^3} +{1315\over1536\o^2} 
                -{3107\over2304\o} +{292849\over6912}
                -{206467\over4608}\o
                +{181740889\over124416}\o^2 \cr
&\phantom{+I_{18}\biggl(}\ 
                -{177573299\over93312}\o^3 \biggr)\biggr]\ . } \eqno(19) $$
Similar expressions hold for the other graphs. 
\par 
By using the above table Eq.s(2,3) are immediately obtained. 
\par 
Graph $1a$ is ultraviolet divergent and requires renormalization. 
From the unrenormalized amplitude we obtain 
$$ \eqalign{  a_e(1a;{\rm not\  ren.}) = 
 &  - {1\over 16 \omega}  + {215\over 24} \Z5  - {1\over 3} \pi^2 \Z3 \cr
 &  - {53\over 2160} \pi^4 
    + 4 \left[\A4 +{1\over24}\left( {\ln^4 2} -\pi^2{\ln^2 2}\right) \right] \cr
 &  - {1229\over 144} \Z3  - {7\over 6} \pi^2 {\ln 2} 
    + {3571\over 2592} \pi^2 + {133\over 864}  \ . \cr }\eqno (20) $$ 
To obtain Eq.(1), one must account for the charge renormalization of the 
inserted 4th order vertex amplitude, to be carried out by means of 
a suitable subtraction constant, which will be indicated here as $Z$. 
For consistency with the previous analytic \gm2 calculations, $Z$ must 
correspond to on mass-shell renormalization. It turns out therefore 
that $Z$ is infrared divergent (this is the only part of the triple-cross 
\gm2 calculation in which infrared divergences appear); again for 
consistency with previous work, the infrared divergence is parametrized 
by giving to the photon a fictitious infinitesimal mass $\lambda$. 
The ultraviolet divergences of the counterterm are still parametrized 
by means of the $n$-dimensional regularization; the result for the 
renormalization counterterm of the inserted 4th order graph reads 
$$ Z = -{1\over8\om} - {11\over24} \pi^2 + 2 - \ll \ .$$
% $$ Z = -{1\over8\om} - {11\over4} \Z2 + 2 - \ll \ .$$
In order to renormalize the contribution of graph $1a$, one must subtract 
from Eq.(20) $ F^{(2)}_2(0) Z $ 
where $ F^{(2)}_2(0) $ is the second order (1 loop) magnetic form factor, 
to be evaluated in $n$-dimensions as well; as $Z$ has a $1/\omega$ 
singularity, use must be done of the value of $ F^{(2)}_2(0) $ evaluated 
up to first order in $ \omega $, 
$$ F^{(2)}_2(0) = {1\over 2} + 2\omega \ . $$ 
By subtracting $ F^{(2)}_2(0) Z $ from Eq.(20), Eq.(1) is recovered. 
%%%%%%%%%%%%%%%%%%%%%%%%%%%%%%%%%%%%%%%%%%%%%%%%%%%%%%%%%%%%%%
\vfill\eject 
\vskip 6 truemm
\parn
{\bf Acknowledgements.} \parn 
As the results presented in this note have been obtained by intensive use 
of the algebra manipulating programs FORM and ASHMEDAI, we want to express 
our gratitude to their authors, J. Vermaseren and M.J. Levine, for their 
help and advice in the early stages of the work. 
% \vfill\eject 
\phantom{.} 
\vskip 2truecm 
\parn
{\bf References}
\def\NCL{{\it Nuovo Cimento Lett.}}

\def\NP{{\it Nucl. Phys. }}
\def\PL{{\it Phys.Lett. }}
\def\PR{{\it Phys.Rev. }}
\def\PRL{{\it Phys.Rev.Lett. }}
\def\ITIM{{\it IEEE Trans. Instrum. Meas.}}
\vskip 10 truemm 
\parn 
\item{[1]}  {S. Laporta and E. Remiddi, \PL {\bf B356}, 390 (1995).} 
\item{[2]}  {M. J. Levine and J. Wright, \PR {\bf D8}, 3171 (1973).} 
\item{[3]}  {T. Kinoshita, \PRL {\bf 75}, 4728 (1995).} 
\item{[4]} M. J. Levine, E. Remiddi and R. Roskies, 
            in {\it Quantum Electrodynamics},
            edited by T. Kinoshita,                                   
            Advanced series on Directions in High Energy Physics, Vol. 7,
            (World Scientific, Singapore, 1990), 162, see p.214-216. 
            Note that the result quoted there in p.215 for the graph $L_2$, 
            which corresponds to Eq.(3) of {R. Barbieri, M. Caffo and 
            E. Remiddi, \NCL {\bf 5}, 769 (1972), is incorrect in that context 
            and should be replaced by Eq.(5) of that same reference, 
            which reads 
            $ -{293\over 72} +{19\over 27} \Z2 +{335\over 144}\Z3 \ .$ 
            We thank B. N. Taylor and P. Mohr for pointing out the oversight.
\item{[5]} {S. Laporta  and E. Remiddi, \PL {\bf B265}, 181 (1991).} 
\item{[6]} {S. Laporta, \PR {\bf D47}, 4793 (1993).} 
\item{[7]} {S. Laporta, \PL {\bf B343}, 421 (1995).} 
\item{[8]} {T. Kinoshita, \ITIM  {} {\bf 44}, 498 (1995).} 
\item{[9]} M. E. Cage {\it et al.}, {\it IEEE Trans. Instrum. Meas.}
            {\bf 38}, 284 (1989) .
\item{[10]} R. S. Van Dick, Jr., P. B. Schwinberg and H. G. Dehmelt,
                    \PRL  {\bf 59}, 26 (1987).
\item{[11]} { see Ref.[4], Sec. 2.1, p. 167.} 
\item{[12]} { K. G. Chetyrkin and F. V. Tkachov, \NP {\bf B192}, 159 (1981); 
            \par F. V. Tkachov, \PL {\bf B100}, 65 (1981). 
            \par We acknowledge also a discussion with D. Broadhurst on 
            this point. } 
\bye